\documentclass[11pt,a4paper,nofootinbib,superscriptaddress]{revtex4}

\pdfoutput=1
\usepackage[utf8]{inputenc}
\usepackage{graphicx,bm}
\usepackage{amssymb,graphicx,epstopdf}
\usepackage{slashed,subfigure}
\usepackage{caption}
\usepackage{xcolor}
\usepackage{amsmath,mathtools}
\usepackage{epstopdf,dcolumn}
\usepackage{soul}
\usepackage{mathtools}
\allowdisplaybreaks
\usepackage[normalem]{ulem}
\usepackage{cancel}
\usepackage{xcolor}
\usepackage[colorlinks=true,linktocpage=true,linkcolor=blue,citecolor=blue]{hyperref}
\usepackage{xcolor}
 \usepackage{braket}
\allowdisplaybreaks
\usepackage{enumitem}

\def\be {\begin{equation}}
\def\ee {\end{equation}}
\def\bea {\begin{eqnarray}}
\def\eea {\end{eqnarray}}
\def\bc {\begin{center}}
	\def\ec {\end{center}}
\def\nn {\nonumber}
\def\eps {\epsilon}
\def\gm {\gamma}

\def\mn {\mu\nu}
\def\al {\alpha}

\def\({\left(}
\def\){\right)}
\def\[{\left[}
\def\]{\right]}
\def\lrarrow{\leftrightarrow}
\def\sp {\shortparallel}
\newcommand \Tr{\operatorname{\text{Tr}}}

\def\sumintf{\sum\!\!\!\!\!\!\!\!\!\!\int\limits}

\def\slashed{\slash\!\!\!\!}

\DeclareGraphicsExtensions{.jpg,.pdf,.eps}

\begin{document}

	\title{Chiral susceptibility in dense thermo-magnetic QCD medium within HTL approximation}
	
	\author{Ritesh Ghosh}
	\email{ritesh.ghosh@saha.ac.in}
	\affiliation{
		Theory Division, Saha Institute of Nuclear Physics, 
		1/AF, Bidhannagar, Kolkata 700064, India}
	\affiliation{
		Homi Bhabha National Institute, Anushaktinagar, 
		Mumbai, Maharashtra 400094, India}
		
		\author{Bithika Karmakar}
	\email{bithika.karmakar@saha.ac.in}
	\affiliation{
		Theory Division, Saha Institute of Nuclear Physics, 
		1/AF, Bidhannagar, Kolkata 700064, India}
	\affiliation{
		Homi Bhabha National Institute, Anushaktinagar, 
		Mumbai, Maharashtra 400094, India}

		\author{Munshi Golam Mustafa}
	\email{munshigolam.mustafa@saha.ac.in}
		\affiliation{
		Theory Division, Saha Institute of Nuclear Physics, 
		1/AF, Bidhannagar, Kolkata 700064, India}
	\affiliation{
		Homi Bhabha National Institute, Anushaktinagar, 
		Mumbai, Maharashtra 400094, India}
	\begin{abstract}
	We have computed the chiral susceptibility in quark-gluon plasma  in  presence of finite chemical potential and weak magnetic field within hard thermal loop approximation. First we construct the massive effective quark propagator in a thermomagnetic medium. Then we obtain completely analytic expression for  the chiral susceptibility  in weak magnetic field approximation. In the absence of magnetic field the thermal chiral susceptibility increases in presence of finite chemical potential. The effect of thermomagnetic correction is found to be very marginal as temperature is the dominant scale in weak field approximation.
	\end{abstract}
	
	\maketitle 
	\newpage
	
\section{Introduction}

	It has been a long standing quest of heavy ion collision community to explore the phase diagram of QCD. Several large scale experiments {\it{e.g.,}} LHC at CERN, RHIC at BNL have been designed and performed for this purpose. Upcoming experiments at FAIR, NICA, JPARC are expected to examine the phase diagram of QCD at high baryon density. Two non-perturbative features of QCD vacuum are confinement and chiral symmetry breaking. With increasing temperature and/or baryon density the QCD vacuum undergoes a phase transition to deconfined and chiral symmetry restored phase. Besides the ongoing experiments, there are several theoretical tools {\it{e.g.,}} lattice calculations~\cite{Aoki:2006we, Bhattacharya:2014ara}, various effective models~\cite{Fukushima:2003fw, Ratti:2005jh}, AdS/QCD correspondence~\cite{Fang:2018axm}, the functional renormalization-group method~\cite{Fischer:2009wc, Braun:2009gm} to study the phase diagram of QCD. Lattice results conclusively demonstrated that the phase transition at vanishing baryon chemical potential is a crossover. The order parameter of chiral symmetry breaking is quark-antiquark condensate which vanishes above the critical temperature in the chiral limit. Chiral susceptibility is the measure of fluctuation of the order parameter. It estimates the response of the chiral condensate with the variation of current quark mass. Measurement of fluctuations is an essential tool to investigate the properties of QCD matter at extreme conditions {\it{e.g.,}} electric charge fluctuation, quark number susceptibility can give insight to the degrees of freedom of the system. Chiral susceptibility has been studied in the framework of lattice QCD~\cite{Karsch:1994hm, Bernard:2004je, Cheng:2006qk, Wu:2006su, Digal:2000ar}, hard thermal loop approximation~\cite{Chakraborty:2002yt}, chiral perturbation theory~\cite{Smilga:1995qf}, NJL model~\cite{Zhuang:1994dw, Sasaki:2006ww}, Dyson-Schwinger equation~\cite{Blaschke:1998mp}, etc. 
	
	Besides, production of magnetic field in non-central heavy ion collisions has added a new dimension to the understanding of QCD matter. This extremely strong magnetic field is created by the spectator particles in a direction perpendicular to the reaction plane. This magnetic field can have detectable consequences like Chiral Magnetic Effect (CME)~\cite{Kharzeev:2013ffa, Fukushima:2012vr}. Other influences of the magnetic field on the QCD matter {\it{viz,}} change in EoS~\cite{Bali:2011qj, Bandyopadhyay:2017cle, Karmakar:2019tdp, Rath:2017fdv}, modification in the transport properties~\cite{Hattori:2017qih, Tuchin:2011jw, Hattori:2016lqx, Kurian:2020qjr, Kurian:2020kct, Kurian:2018qwb, Kurian:2018dbn}, dilepton production rate~\cite{Tuchin:2013ie, Tuchin:2013bda, Bandyopadhyay:2016fyd, Ghosh:2018xhh, Das:2019nzv}, heavy quark potential~\cite{Hasan:2017fmf, Singh:2017nfa}, damping rate of photon~\cite{Ghosh:2019kmf} have been studied by different groups of heavy ion collision community. The magnetic field can affect the dynamical chiral symmetry breaking. Some studies suggest that the chiral condensate increases in the presence of magnetic field. It is argued that in case of neutral spin-zero pair of fermion and antifermion, magnetic moments of both point along same direction. As a result, both magnetic moments can align along the magnetic field direction without creating any frustration in the fermion-antifermion pair~\cite{Shovkovy:2012zn}. This effect is linked to the increase in the phase transition temperature which is known as Magnetic Catalysis. However, several lattice studies~\cite{Bali:2012zg} have found the opposite nature {\it{i.e.,}} the decrease in phase transition temperature at least for small magnetic fields. This effect has been named as Inverse Magnetic Catalysis. Also these studies revealed that the change in the chiral condensate strongly depends on the temperature and the quark mass. Recently, the chiral susceptibility was calculated using NJL model in Ref.~\cite{Das:2019crc} in the presence of chiral chemical potential and non-zero magnetic field. The magnetic field breaks the flavor symmetry. Hence two distinct peaks of chiral susceptibility for u and d quark have been observed at large magnetic field.

The strong magnetic field produced in heavy ion collision sharply decays with time~\cite{Bzdak:2012fr,McLerran:2013hla}. However, some studies~\cite{Tuchin:2013bda, Tuchin:2013ie} have shown that the presence of finite conductivity can make the strong magnetic field survive for long time. 
	The QCD matter cools down after the collision and undergoes the chiral phase transition at around 160 MeV temperature. In this region the effect of weak magnetic field is particularly important. In this paper, we consider the magnetic field to be small and use the scale hierarchy $\sqrt{|q_fB|} < gT < T$. In Ref.~\cite{Chakraborty:2002yt} the chiral susceptibility was computed with zero chemical potential within hard thermal loop (HTL) approximation. In this paper we, considering recently obtained effective quark propagator in the presence of weak magnetic field~\cite{Das:2017vfh}, determine the chiral susceptibility with finite chemical potential in the QCD medium using HTL approximation.
	
The paper is organized as follows. In Sec.~\ref{definition} we describe the static chiral susceptibility. We obtain the general structure of fermion self energy in presence of weak magnetic field and compute the effective propagator in Sec.~\ref{gs}. The free chiral susceptibility is calculated in Sec.~\ref{cs_free}. We compute the HTL chiral susceptibility within weak magnetic field approximation in Sec.~\ref{cs_htl}. The results are  described in Sec.~\ref{result}. Finally, we summarize in Sec.~\ref{summary}.

\section{Definition}
\label{definition}
The chiral condensate is defined as 
\bea
\braket{\bar q q} &=& \frac{\Tr[\bar q q\, e^{-\beta \rm{H}}]}{\Tr[e^{-\beta \rm{H}}]}=\frac{\partial \Omega}{\partial m_f},
\eea
where H is the Hamiltonian of the system. $\Omega=-\frac{T}{V}\ln Z$ is the thermodynamic potential where Z is the partition function of a quark-antiquark gas.
Quark condensate also can be written using quark propagator as
\bea
\braket{\bar q q} &=& -N_c N_f \sumintf_{\{P\}}\Tr\bigg[S(P)\bigg],\label{condensate}
\eea
where $N_c$ and $N_f$ are the numbers of quark colors and flavors respectively. 
Susceptibility is the measure of the response of a system to small external force. Chiral susceptibility measures the response of chiral condensate to infinitesimal change of current quark mass $m_f$ as
\bea
\chi_c &=& -\frac{\partial \braket{\bar q q}}{\partial m_f}\bigg|_{m_f=0}.\label{cs_def}
\eea


\section{General structure of Fermionic two point function}
\label{gs}
Recently covariant structure of fermion self-energy has been constructed in the presence of temperature and magnetic field in Ref.~\cite{Das:2017vfh}. 
General covariant structure of fermion self energy in a weak thermomagnetic field can be written as
\bea
\Sigma(P)&=& -a \,\slashed{P}-b\, \slashed u -c'\, \gm_5\, \slashed u- d'\, \gm_5\, \slashed n,
\label{self-energy}
\eea
where $u_\mu$ is four velocity of fluid. The direction of magnetic field $n_\mu$ can be written in terms of electromagnetic field tensor $F^{\mn}$ or its dual $\tilde F^{\mn}$ and fluid velocity $u_\mu$ as
\bea
n_\mu \equiv \frac{1}{2B} \epsilon_{\mu\nu\rho\lambda}\, u^\nu F^{\rho\lambda} 
= \frac{1}{B}u^\nu {\tilde F}_{\mu\nu}. \label{bmu}
\eea
For simplicity we have chosen the fluid rest frame and the magnetic field along $z$-direction as
\bea
u^{\mu}&=& (1,0,0,0),\\
n_\mu &=& (0,0,0,1).
\eea 
The  self-energy structure functions $a$, $b$, $c'$ and $d'$ in Eq.~\eqref{self-energy}  can be calculated using Eqs.~\eqref{a_calc}, \eqref{b_calc}, \eqref{c'_calc} and \eqref{d'_calc} in Appendix~\ref{appendix_A}. The structure functions in the presence of weak magnetic field are calculated upto $\mathcal O(q_fB)$ for zero quark chemical potential in Ref.~\cite{Das:2017vfh}. The calculations are generalized for finite quark chemical potential in Ref.~\cite{Bandyopadhyay:2017cle}. Here, we compute the structure functions upto $\mathcal O(q_f B)^2$ in presence of chemical potential in Appendix~\ref{appendix_A}.

Following the Dyson-Schwinger equation, the effective inverse propagator of massive fermion can be written as 
\bea
S_{\text {eff}}^{-1} &=&\slashed P - m_f \mathbb{I}- \Sigma.
\eea
Using Eq.~\eqref{self-energy}  the structure of the inverse propagator of massive fermion in thermomagnetic medium can be written as 
\bea
S_{\text {eff}}^{-1}(P)&=& (1+a)\, \slashed{P}+b \,\slashed u +c'\, \gm_5\, \slashed u+ d' \, \gm_5\, \slashed n-m_f\, \mathbb{I}.
\label{eff_inv}
\eea
To compute the  chiral susceptibility in the presence of weak magnetic field one requires the effective fermion propagator as given in Eqs.~\eqref{condensate} and \eqref{cs_def}. So we need to invert Eq.~\eqref{eff_inv} to get the effective fermion propagator. For massless case, it is very easy to invert the effective inverse propagator to obtain the general structure of effective propagator in terms of  $\slashed P, \,\slashed u$, $\gm_5\,\slashed u$ and \,$ \gm_5\, \slashed n $. To get the structure of the effective propagator in the massive case involving the Dirac matrices:  $\slashed P,\,\slashed u$, $\gm_5\,\slashed u,\, \gm_5\, \slashed n $ and $\mathbb{I}$, we adopt the following trick used in Ref.~\cite{Das:2019ehv}.

Let us assume that we need to find the inverse of a matrix $M$. Now we need to choose a matrix $R$ and multiply it with $M$ to get a matrix $U$ as
\bea
U=M R.
\label{U}
\eea
Now we can write inverse of the matrix $M$ as
\bea
M^{-1}= R U^{-1}.
\label{Minv}
\eea
In our case we need to find the inverse of the matrix $S_\text{eff}^{-1}$. Now it is essential to choose $R$ in such a way that we get $U$ in Eq.~\eqref{U} in a very simple form. Then it would be easy to invert the matrix $U$ and to find the inverse of our desired matrix $S_\text{eff}^{-1}$.

Thus we choose $R$ as
\bea
R&=& (1+a)\, \slashed{P}+b\, \slashed u -c' \gm_5\, \slashed u- d' \gm_5\, \slashed n - m_f  \mathbb{I}.
\eea
From Eqs.~\eqref{eff_inv} and \eqref{U}, we have
\bea
U=S_\text{eff}^{-1} R &=& \alpha \, \slashed P +\beta \, \slashed u +\delta\, \gm_5 + \lambda\,  \mathbb{I},
\eea
where
\bea
\alpha &=& -2 (1+a) \,m_f, \nn\\
\beta &=& -2 b\, m_f,\nn\\
\delta &=& 2 ((1+a) c' p_0+b c' + (1+a) d' p_3), \nn\\
\lambda &=& (1+a)^2 P^2+b^2+ c'^2-d'^2+m_f^2+ 2 (1+a) b p_0.
\eea
We can now easily invert the matrix U to get
\bea
U^{-1}&=&\frac{1}{N^2}(\alpha \, \slashed P +\beta \, \slashed u  +\delta\, \gm_5  - \lambda\,  \mathbb{I}),
\eea
where 
\bea
N^2 &=& \alpha^2 P^2 + 2 \alpha \beta p_0+ \beta^2 +\delta^2-\lambda^2 .
\eea
Following Eq.~\eqref{Minv}, we get the effective fermion propagator $S_{\text {eff}} $ as 
\bea
S_{\text {eff}} &=&R U^{-1}= \Big( (1+a)\, \slashed{P}+b\, \slashed u -c' \gm_5\, \slashed u- d' \gm_5\, \slashed n-m_f \,  \mathbb{I} \Big) \frac{ \alpha \, \slashed P +\beta \, \slashed u  +\delta\, \gm_5  - \lambda\,  \mathbb{I}}{ \alpha^2 P^2 + 2 \alpha \beta p_0+ \beta^2 +\delta^2-\lambda^2}.\label{eff_prop}
\eea
The dispersion relation for massive fermion in weakly magnetized thermal medium can be obtained from the denominator of the effective propagator by setting it to zero. 

\section{Chiral susceptibility for free fermion in the presence of weak magnetic field}
\label{cs_free}
We consider weakly magnetized QCD medium. In the weak magnetic field limit, we work with the scale hierarchy, $\sqrt{|q_f B|}<m_{\mathrm{th}}\sim gT<T$. Now treating  $q_f B$ as perturbation, the Schwinger propagator for a fermion in presence of weak magnetic field can be expanded and written up to $\mathcal{O}[(q_f B)^2]$ as~\cite{Chyi:1999fc}
\bea
S(K)&=& \frac{\slashed{K}+m_f}{K^2-m_f^2}+ i\gm^1\gm^2\frac{\slashed{K_{\sp}}+m_f}{(K^2-m_f^2)^2}(q_f B)+  \ 2  \left[\frac{\left\{(K\cdot u)\,\slashed{u}-(K\cdot n)\,\slashed{n}\right\} 
	-\slashed{K}}{(K^2-m_f^2)^3}-\frac{k_\perp^2(\slashed{K}+m_f)}{(K^2-m_f^2)^4}\right](q_f B)^2\nn\\
&+& \mathcal{O}\left[(q_f B)^3\right]\nn\\
&=& S_0+S_1+S_2+\mathcal{O}[(q_f B)^3].\label{schwinger_prop}
\eea

We can write chiral condensate for free fermion in weak magnetic field upto $\mathcal{O}[(q_f B)^2]$ from Eq.~\eqref{schwinger_prop} as
\bea
\braket{\bar q q}_f &=& -N_c N_f \, \sumintf_{\{P\}}\Tr\bigg[S_0(P)+S_1(P)+S_2(P)\bigg],\nn\\
&=&  -4 m_f\,N_c N_f \,\sumintf_{\{P\}} \bigg[ \frac{1}{P^2-m_f^2}-2\,(q_f B)^2\frac{p_\perp^2}{(P^2-m_f^2)^4} \bigg].
\label{cs_free_B}
\eea
Using the definition in Eq.~\eqref{cs_def} the chiral susceptibility for free fermion in weak magnetic field can be calculated as
\bea
\chi_c=  -\frac{\partial \braket{\bar q q}}{\partial m_f}\bigg|_{m_f=0}&=&4 N_c N_f \,\sumintf_{\{P\}} \bigg[ \frac{1}{P^2}-2\,(q_f B)^2\frac{p_\perp^2}{(P^2)^4} \bigg]\nn\\
&=& \frac{N_c N_f}{6} T^2 \bigg[1+12\hat\mu^2-(q_f B)^2 \frac{\gimel(z)}{16 \pi^4 T^4}\bigg],
\eea

where $\hat \mu=\mu/2\pi T$. $\mu$ is the quark chemical potential. The sum-integrals are calculated in Appendix~\ref{sum_int} and $\gimel(z)$ is given in Eq.~\eqref{alpha}.


\section{HTL chiral Susceptibility in presence of weak magnetic field}
\label{cs_htl}
Using the effective quark propagator in Eq.~\eqref{eff_prop} chiral condensate $\braket{\bar q q}$ takes the form,
\bea
\braket{\bar q q}&=& - N_c\, N_f\,  \sumintf_{\{P\}}\, \Tr[S_{\text{eff}}(P)]\nn\\
&=&  4 m_f N_c\, N_f\,  \sumintf_{\{P\}} \frac{(1+a)^2 \,P^2+ 2 (1+a)\,b\, p_0+b^2+d'^2-c'^2-m_f^2}{ \alpha^2 P^2 + 2 \alpha \beta p_0+ \beta^2 +\delta^2-\lambda^2} .\label{htl_condensate}
\eea

Chiral susceptibility in the massless limit can be calculated from Eq.~\eqref{htl_condensate} as
\bea
\chi_c &=& -\frac{\partial \braket{\bar q q}}{\partial m_f}\bigg|_{m_f=0}\nn\\
&=& -4 N_c\, N_f\,  \sumintf_{\{P\}} \frac{(1+a)^2\,P^2 + 2 (1+a)\,b \, p_0+b^2+d'^2-c'^2}{4 \Big[(1+a) c' \,p_0+b c' + (1+a) d' \,p_3\Big]^2-\Big[(1+a)^2 P^2+b^2+ c'^2-d'^2+ 2 (1+a)\, b \,p_0\Big]^2},\nn\\
\label{cs}
\eea
where the expressions of various structure functions are obtained in Appendix~\ref{appendix_A}.
Now we expand the expression in Eq.~\eqref{cs} in the series of coupling constant $g$ and keep upto $\mathcal O(g)^4$ as
\bea
\chi_c &=& 4 N_c\, N_f\,  \sumintf_{\{P\}}\, \bigg\{\frac{1}{P^2}+ 2 m_{\text{th}}^2 \frac{1}{P^4}+m_{\text{th}}^4\bigg(\frac{4}{P^6}+\frac{1}{p^2\, P^4}-\frac{2}{p^2\,P^4}\mathcal{T}_p+\frac{1}{p^2 \,p_0^2 \,P^2}\mathcal{T}_p^2\bigg)\nn\\
&-&m'^2_{\text{eff}}\bigg(\frac{2}{3 P^4}-\frac{2 p_3^2}{p^2 P^4}-\frac{2}{P^4}\mathcal{T}_p+\frac{2 p_3^2}{p^2 P^4}\mathcal{T}_p\bigg) \nn\\
&-& m_{\text{eff}}^4 \bigg(\frac{4p_3^2}{p^2 P^6}+\frac{p_3^2}{p^4P^4}-\frac{2p_3^2 }{p^4 P^4}\mathcal{T}_p+\frac{3}{p_0^2 P^4}\mathcal{T}_p^2+\frac{p_3^2 }{p^4 P^4}\mathcal{T}_p^2-\frac{4p_3^2}{p^2 p_0^2 P^4}\mathcal{T}_p^2\bigg)\bigg\}\label{chi_exp},
\eea
where
\bea
m^2_\text{th}&=& \frac{g^2C_F T^2}{8}\Big( 1+4\hat \mu^2\Big),\nn\\
m'^2_\text{eff}&=& \frac{g^2 C_F (q_f B)^2 T}{32 \,\pi \,m_f^3},\nn\\
m^2_\text{eff}&=&  4 g^2\,C_F\,\frac{q_fB}{16\pi^2}\bigg[-\frac{1}{4}\aleph(z)-\frac{\pi T}{2 m_f}-\frac{\gamma_E}{2}\bigg],
\eea
with $\aleph(z)$  defined in Eq.~\eqref{aleph} and $C_F=(N_c^2-1)/2 N_c$  is QCD Casimir factor.
Using the sum-integrals listed in Appendix~\ref{sum_int} we find the expression for the  chiral susceptibility as
\bea
\chi_c&=&\frac{N_c N_f}{6}T^2 \bigg[ 1+12\hat\mu^2+\frac{3}{\pi^2}\bigg(\frac{\Lambda}{4\pi T} \bigg)^{2\eps}\bigg(\frac{1}{\eps}-\aleph(z) \bigg)\frac{m_\text{th}^2}{T^2}\nn\\
&+&\frac{1}{ \pi^2}\bigg(\frac{\Lambda}{4\pi T} \bigg)^{2\eps}\bigg(\frac{1}{\eps}+\frac{4 }{3}-\aleph(z)\bigg)\frac{m'^2_{\text{eff}}}{T^2}
+\frac{\gimel(z)}{32\pi^4}\Big( \pi^2-6\Big)\frac{m_\text{th}^4}{T^4} -\frac{\gimel(z)}{24\pi^4} \Big( \pi^2-6\Big)\frac{m_\text{eff}^4}{T^4}\bigg].
\eea
We note that the logarithmic divergence comes from the thermal part. A new divergence appears in presence of the magnetic field. We renormalize the chiral susceptibility within $\overline{MS}$ renormalization scheme using the following counter term
\bea
\Delta \chi_c^{counter}&=&-\frac{N_c N_f}{6 \pi^2 \eps}\left(3 m_\text{th}^2+m'^2_{\text{eff}}\right ).
\eea
The renormalized chiral susceptibility is given as
\bea
\chi_c&=&\frac{N_c N_f}{6}T^2 \bigg[ 1+12\hat\mu^2+\frac{3}{\pi^2}\bigg(2\ln \hat \Lambda -2\ln 2-\aleph(z)  \bigg)\frac{m_\text{th}^2}{T^2}\nn\\
&+&\frac{1}{3 \pi^2}\bigg(4-3 \aleph(z)+6 \ln \hat \Lambda-6 \ln 2\bigg)\frac{m'^2_{\text{eff}}}{T^2}+\frac{\gimel(z)}{32\pi^4}\Big( \pi^2-6\Big)\frac{m_\text{th}^4}{T^4} -\frac{\gimel(z)}{24\pi^4} \Big( \pi^2-6\Big)\frac{m_\text{eff}^4}{T^4}\bigg],
\label{cs_renormalised}
\eea
with $\hat \Lambda=\Lambda/2\pi T$ and $\hat \mu=\mu /2\pi T$. The obtained result  is completely analytic in presence of chemical potential and weak magnetic field.
Here we note that the Eq.~\eqref{cs_renormalised} consists of $\mathcal O[(q_f B)^0]$ and $\mathcal O[(q_f B)^2]$ terms. The $\mathcal  O[(q_fB)^0]$ reproduces the thermal chiral susceptibility without chemical potential obtained in Ref.~\cite{Chakraborty:2002yt}. The $\mathcal O[(q_f B)^2]$ terms are the thermomagnetic correction to the thermal chiral susceptibility.

\section{Results}
\label{result}
We  consider  magnetic field dependent running coupling \cite{Ayala:2018wux} as
\bea
\alpha_s (\Lambda^2,|eB|)=\frac{\alpha_s(\Lambda^2)}{1+b_1 \alpha_s(\Lambda^2) \ln \left(\frac{\Lambda^2}{\Lambda^2+|eB|}\right)},
\eea 
where the one loop running coupling at renormalization scale reads as
\bea
\alpha_s(\Lambda^2)=\frac{1}{b_1 \ln\left(\Lambda^2/\Lambda^2_{\overline{\rm{MS}}}\right)},
\eea
with $b_1=\frac{11 N_c-2 N_f}{12 \pi}$, $\Lambda_{\overline{\rm{MS}}}=204$ MeV requiring $\al_s=0.326$ at 1.5 GeV \cite{Beringer:1900zz}. We choose the renormalization scale as $\Lambda=2\pi \sqrt{T^2+\mu^2/\pi^2}$. The following results are shown considering two light quark flavors $u$ and $d$.

\begin{center}
	\begin{figure}[tbh]
		\begin{center}
			\includegraphics[scale=0.57]{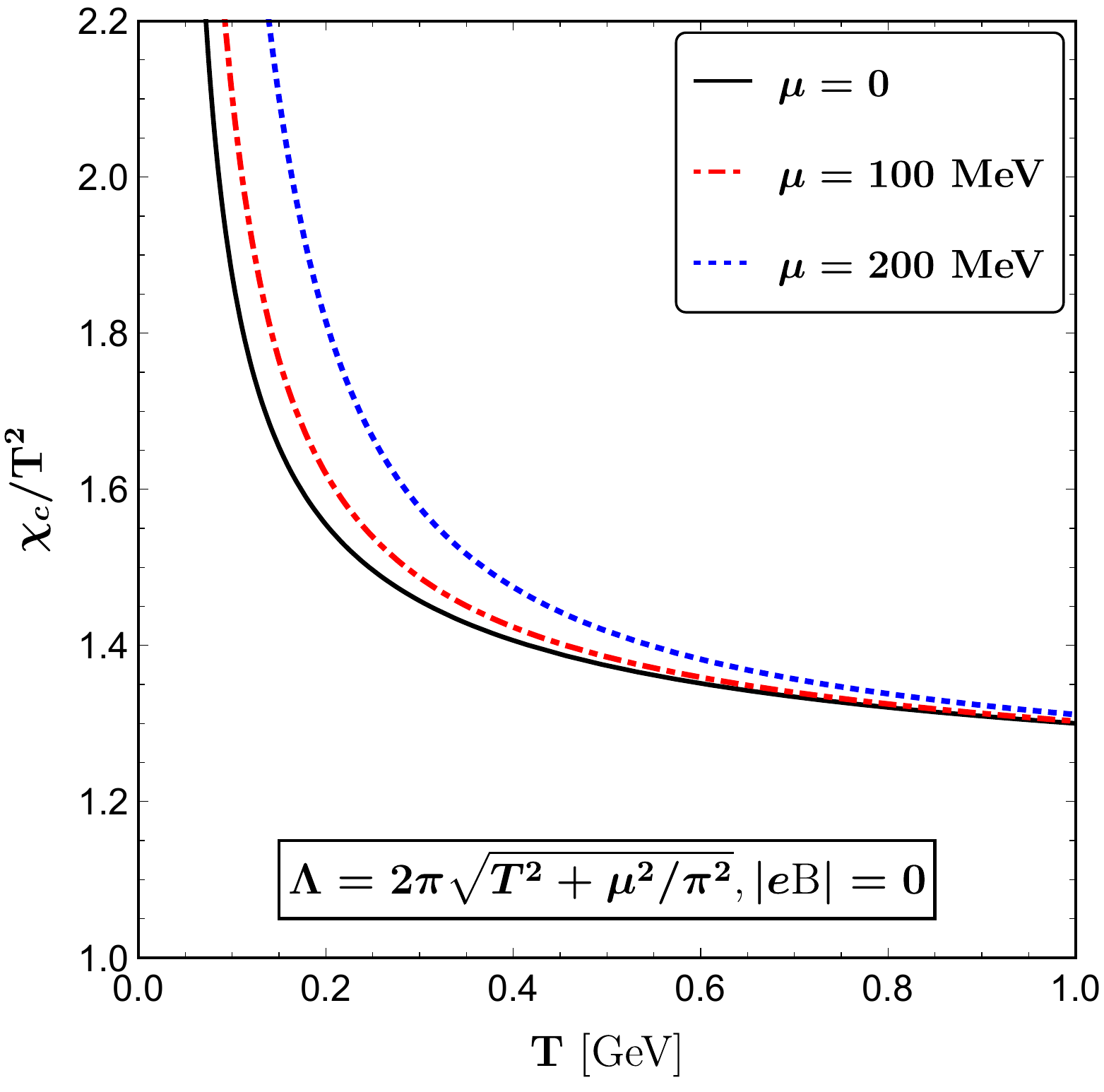}
			\caption{ Variation of chiral susceptibility scaled with $T^2$ as a function of temperature for  chemical potential 0, 100 and 200 MeV with  zero magnetic field.}
			\label{cvT_B0}
		\end{center}
	\end{figure}
\end{center}

In Fig.~\ref{cvT_B0} the chiral susceptibility scaled with temperature squared is plotted  with temperature  in absence of magnetic field for zero and non zero quark chemical potential. The effect of quark chemical potential is prominent in the low temperature region as can be seen from the figure. Similar plot for thermal QCD medium and zero chemical potential was obtained in Ref.~\cite{Chakraborty:2002yt}. For low temperature the chiral susceptibility increases rapidly for both zero and non zero chemical potential. Here we note that the increase of the chiral susceptibility in the low temperature region does not indicate the chiral phase transition. It is due to the temperature dependence of the coupling constant and for the choice of the renormalization scale~\cite{Chakraborty:2002yt}. At very high temperature the chiral susceptibility reaches the free value asymptotically.

\begin{center}
	\begin{figure}[tbh]
		\begin{center}
			\includegraphics[scale=0.5]{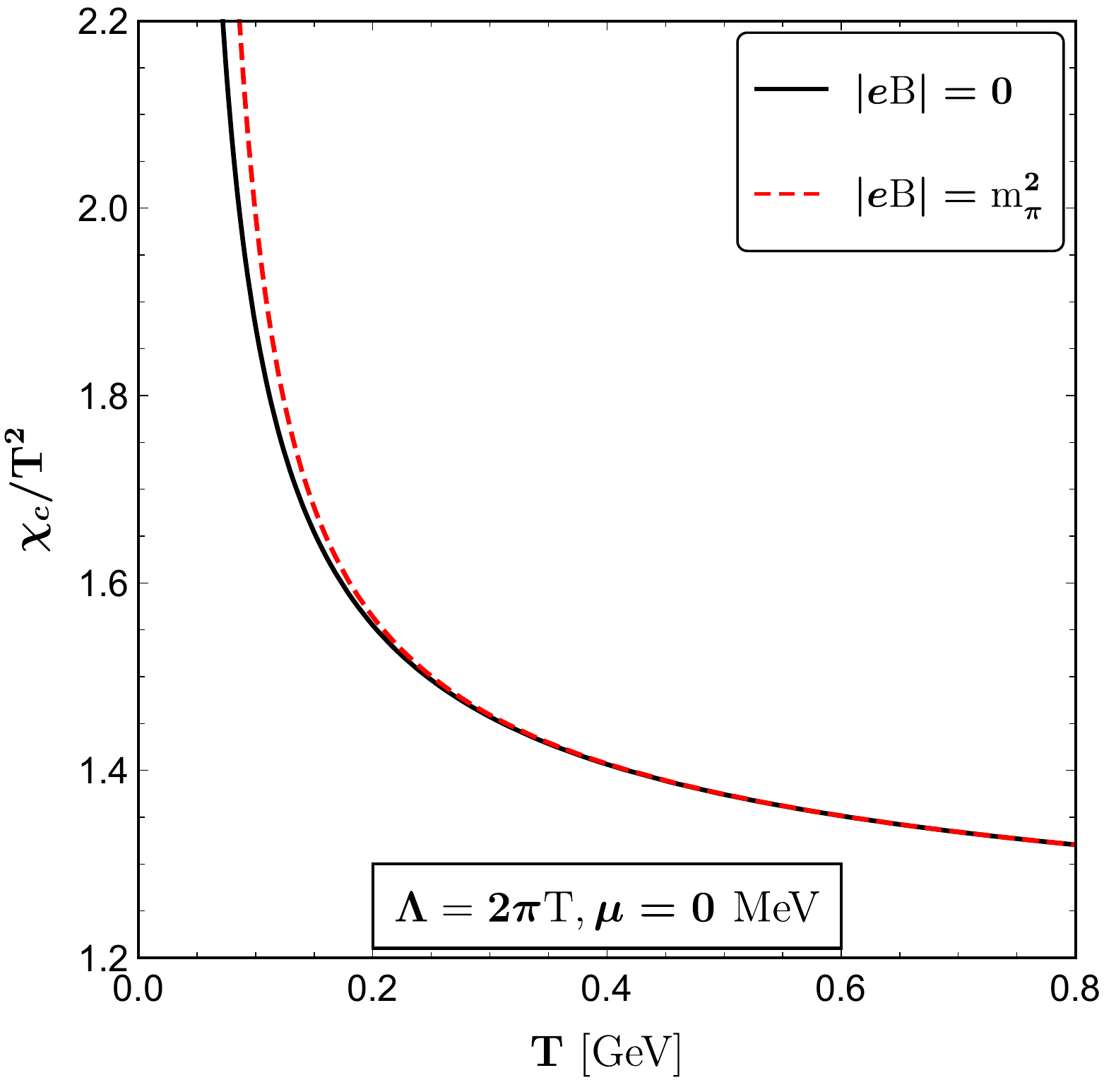}
			\includegraphics[scale=0.5]{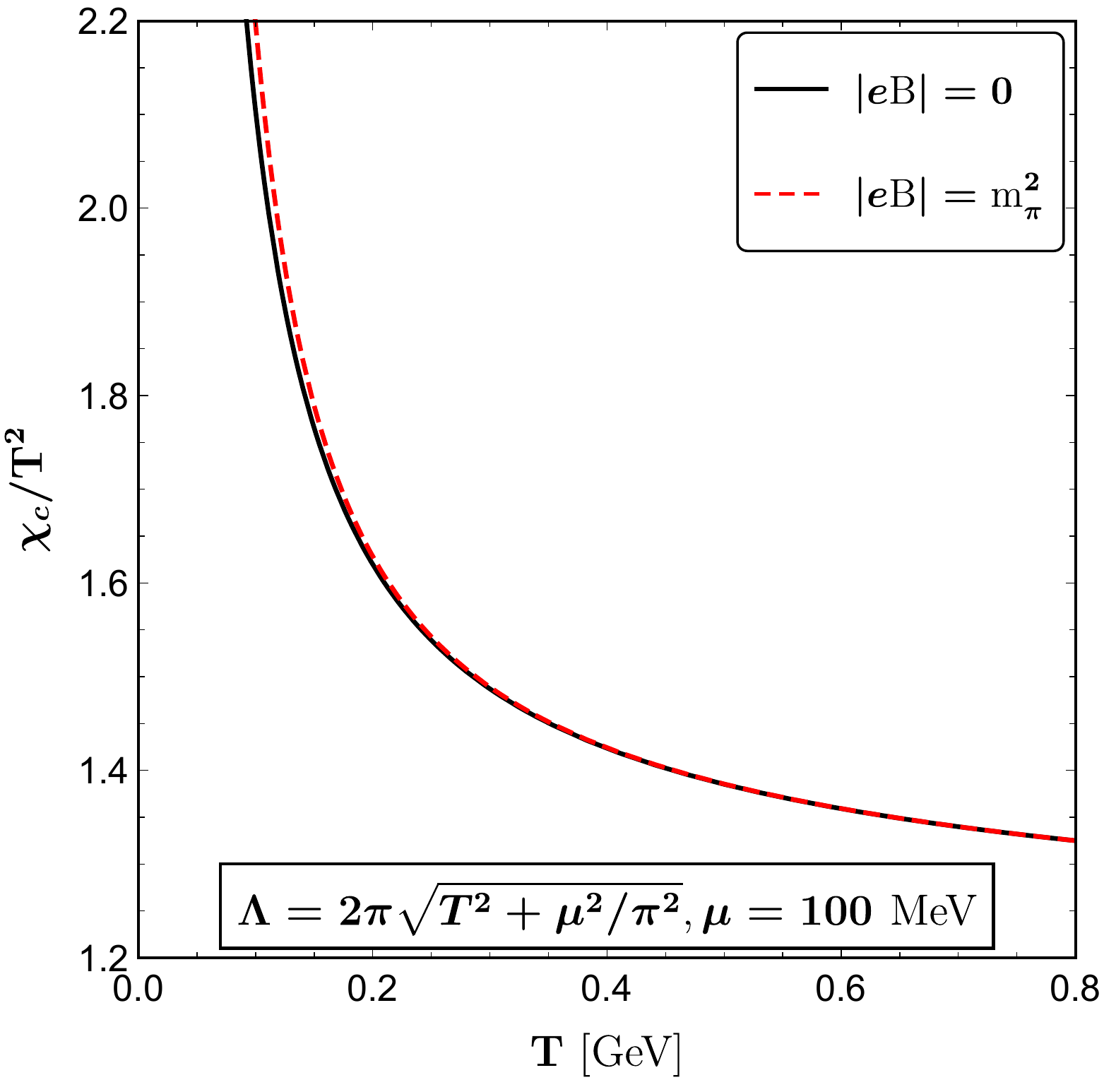}
		  \end{center}
			\caption{ Variation of chiral susceptibility scaled with $T^2$ as a function of $T$  for magnetic field strength $|eB|=0, m_\pi^2$  with $\mu=0$ MeV (left) and $\mu= 100$ MeV (right).}
			\label{cvT_mu}
	\end{figure}
\end{center}
  The variation of chiral susceptibility  scaled with temperature squared for zero and finite magnetic field is plotted with temperature in Fig.~\ref{cvT_mu}. In the left panel of Fig.~\ref{cvT_mu} we have shown the effect of weak magnetic field on the chiral susceptibility for zero quark chemical potential, whereas, the same for finite quark chemical potential is shown in the right panel. In presence of magnetic field chiral susceptibility is slightly increased than that of thermal medium in the low temperature region .Since we are working in weak magnetic field limit, the increase in susceptibility due to magnetic field is small. As temperature increases the effect of magnetic field reduces as temperature becomes the dominant scale. 
  
   It should be noted that the scale hierarchy of weakly magnetized medium is satisfied for around $T > 0.14 $ $GeV$ as we have considered $|eB|=m_\pi^2=0.14^2$ $GeV^2$ in Fig.2. Thus the weak field and HTL approximations are valid at high temperature.

\begin{center}
	\begin{figure}[tbh!]
		\begin{center}
			\includegraphics[scale=0.5]{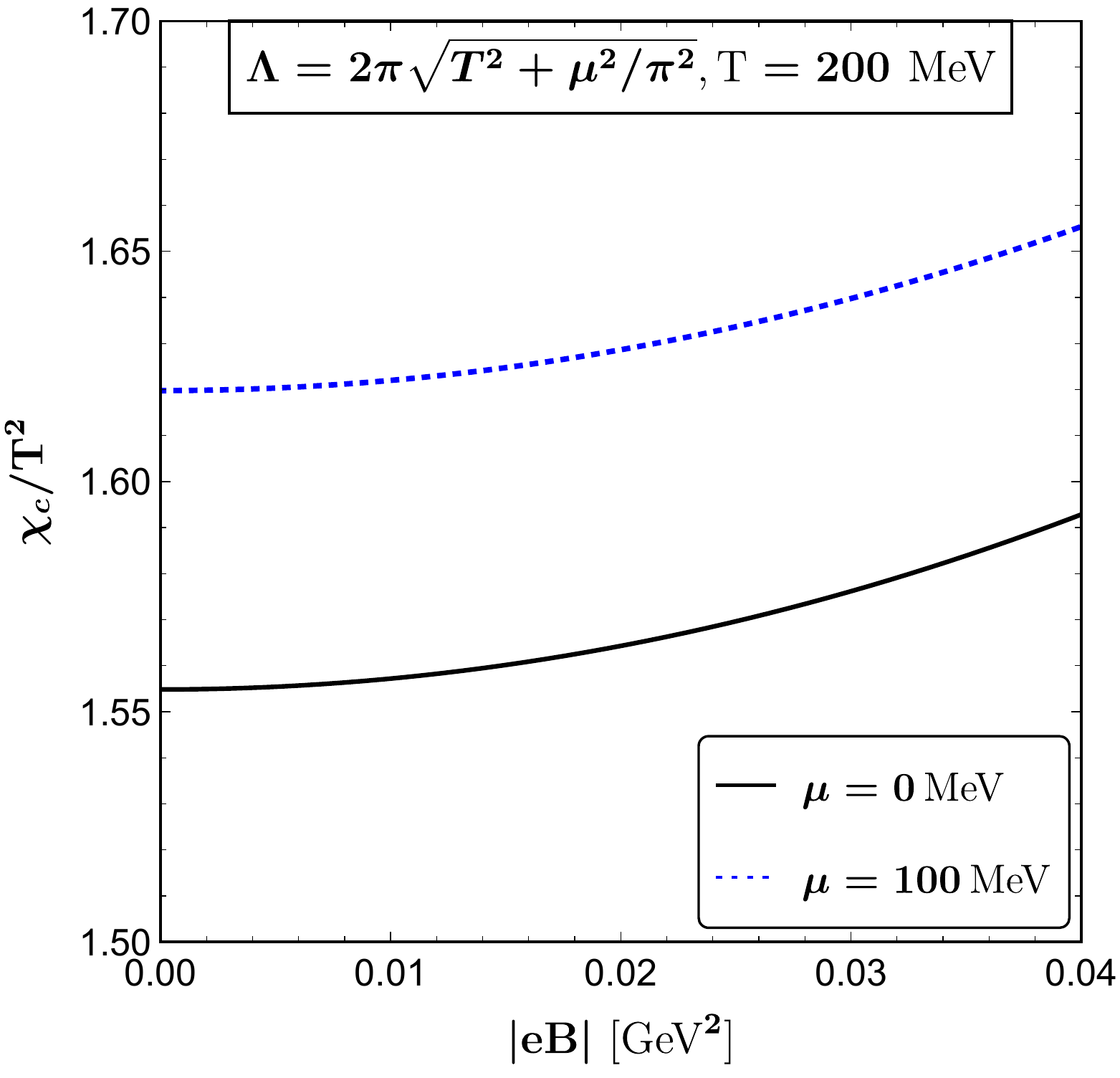}
			\caption{Scaled chiral susceptibility is plotted as a function of magnetic field strength $|eB|$ for temperature $T= 0.2$ MeV and $\mu=0$ MeV   and 100 MeV.}
			\label{cvB_mu}
		\end{center}
	\end{figure}
\end{center}
The effect of magnetic field can be seen clearly from Fig.~\ref{cvB_mu} where the variation of the scaled chiral susceptibility is shown with magnetic field for fixed temperature $T=200$ MeV. Here we notice the slow increase in the chiral susceptibility with increasing magnetic field for both with and without chemical potential.

\begin{center}
	\begin{figure}[tbh!]
		\begin{center}
			\includegraphics[scale=0.5]{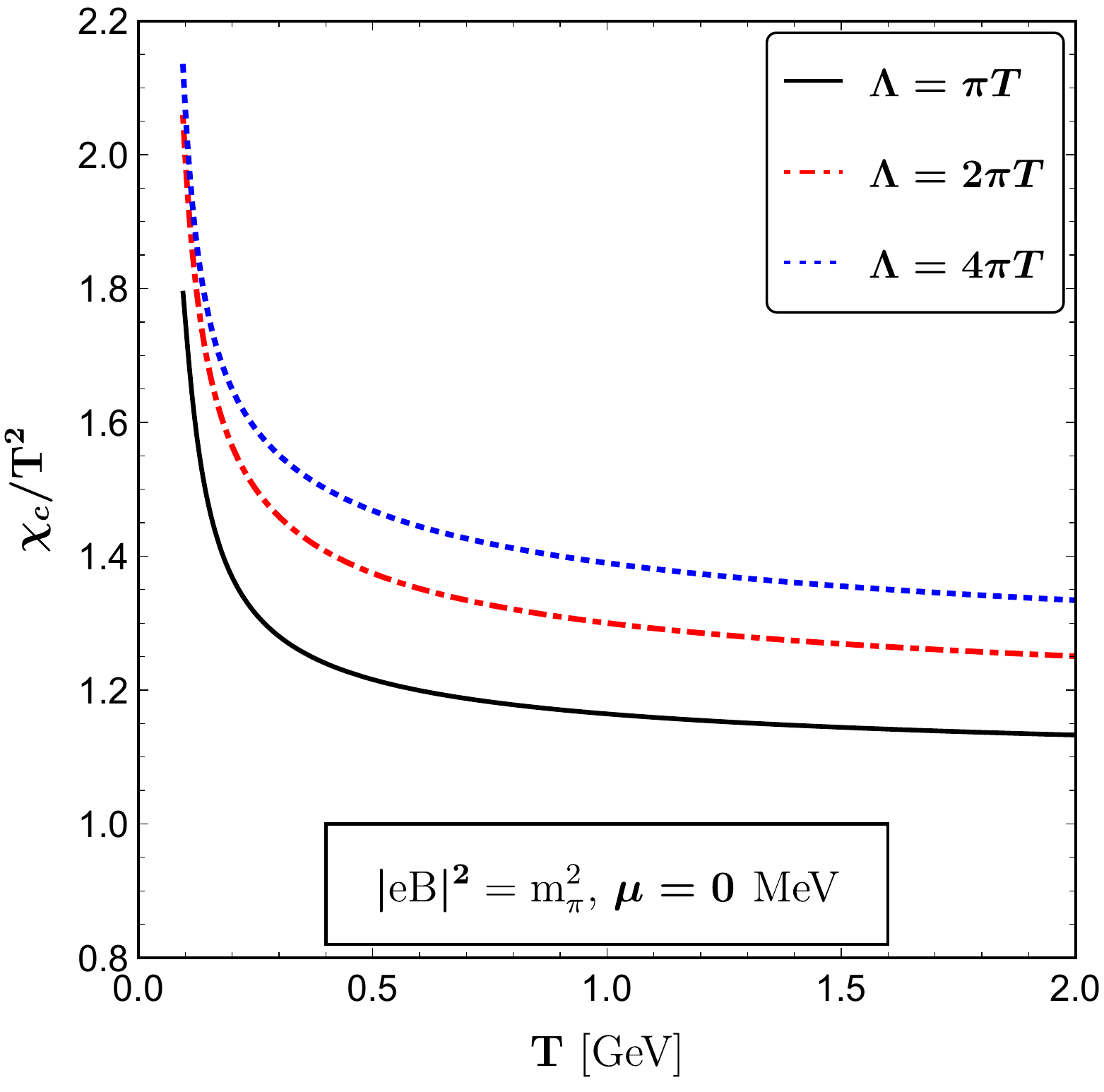}
				\includegraphics[scale=0.5]{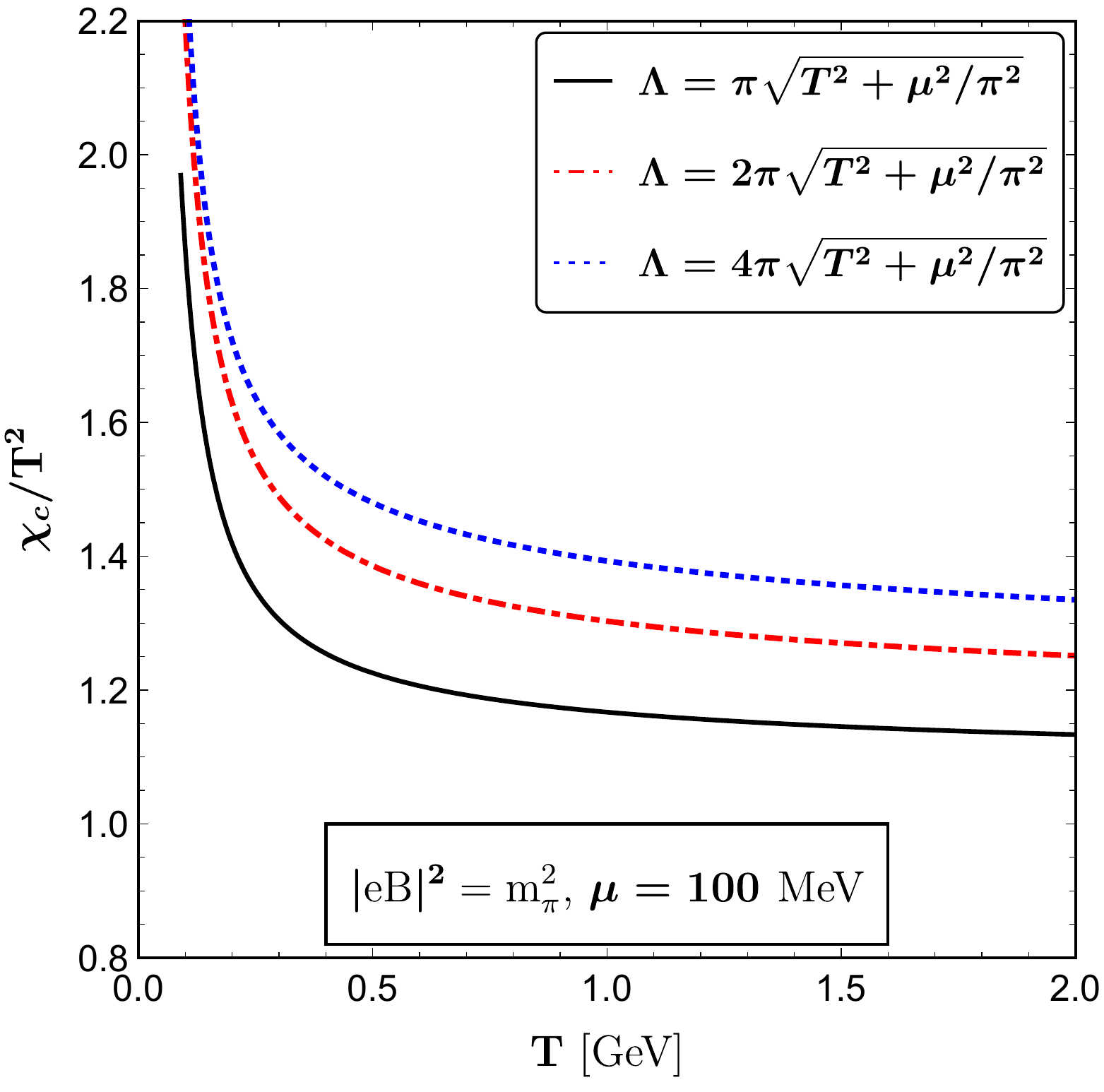}
			\caption{  Variation of chiral susceptibility scaled with $T^2$ as a function of temperature  for magnetic field strength $ m_\pi^2$ for different values of renormalization scale.   }
			\label{cvT_lambda}
		\end{center}
	\end{figure}
\end{center}
In Fig.~\ref{cvT_lambda} the sensitivity of the chiral susceptibility with renormalization scale is shown in the presence of a constant
weak magnetic field. Here chiral susceptibility scaled with $T^2$ is plotted with temperature for zero(left panel) and finite (right panel) chemical potential by varying renormalization scale $\Lambda$
by factor 2 around its central value $2\pi \sqrt{T^2 +\mu^2/\pi^2}$. 

Here we note that HTL approximation is valid above the phase transition temperature where the scale hierarchy  $\sqrt{|q_fB|} < gT < T$ is maintained. We have shown the plots of chiral susceptibilities at low temperature just to show the steep increase in the plots. 
\section{Summary}
\label{summary}
We have investigated the effect of magnectic field on the chiral susceptibility of quark-gluon plasma within HTL approximation in presence of finite chemical potential. The general structure of effective massive fermion propagator is constructed for thermo-magnetic medium. Then the self-energy structure functions upto $\mathcal{O}[(q_f B)^2]$ in presence of chemical potential have been calculated in the weak magnetic field regime using the scale hierarchy $\sqrt{|q_f B|}<gT< T$.  Quark condensate is computed using effective  quark propagator in presence of magnetic field. Finally we obtain completely analytic  expression for chiral susceptibility in hot and dense weakly magnetized QCD medium. We have subtracted the UV divergence via $\overline{MS}$ renormalization scheme. It is found that the  chiral susceptibility is increased due to the presence of chemical potential as well as the background magnetic field. At high temperature the effect of magnetic field on chiral susceptibility becomes feeble.

\section*{ACKNOWLEDGMENTS}
RG is funded by University Grants Commission (UGC). BK and MGM are funded by Department of Atomic Energy (DAE), India via the 
project TPAES .

\appendix

\section{Structure functions}
\label{appendix_A}
The general form of the various structure functions can be written from Eq.~\eqref{self-energy} as
\bea
a&=& \frac{1}{4}\frac{\Tr(\Sigma \, \slashed P)-(P\cdot u) \Tr(\Sigma\, \slashed u)}{(P\cdot u)^2-P^2},\label{a_calc}\\
b&=& \frac{1}{4}\frac{-(P\cdot u)\Tr(\Sigma \, \slashed P)+P^2 \Tr(\Sigma\, \slashed u)}{(P\cdot u)^2-P^2},\label{b_calc}\\
c'&=&-\frac{1}{4} \Tr(\,\slashed u \,\Sigma \gm_5),\label{c'_calc}\\
d'&=&\frac{1}{4} \Tr(\,\slashed n \, \Sigma \gm_5)\label{d'_calc}.
\eea
The structure functions in the presence of magnetic field depends on three Lorentz scalars
\bea
p_0&=& P^\mu u_\mu,\\
p_3&=&P^\mu n_\mu=p_z,\\
p_\perp&=&[-(P^\mu P_\mu)^2+( P^\mu u_\mu)^2-( P^\mu n_\mu)^2]^{1/2}=(p_1^2+p_2^2)^{1/2}.
\eea

\begin{center}
	\begin{figure}[tbh]
		\begin{center}
			\includegraphics[scale=0.4]{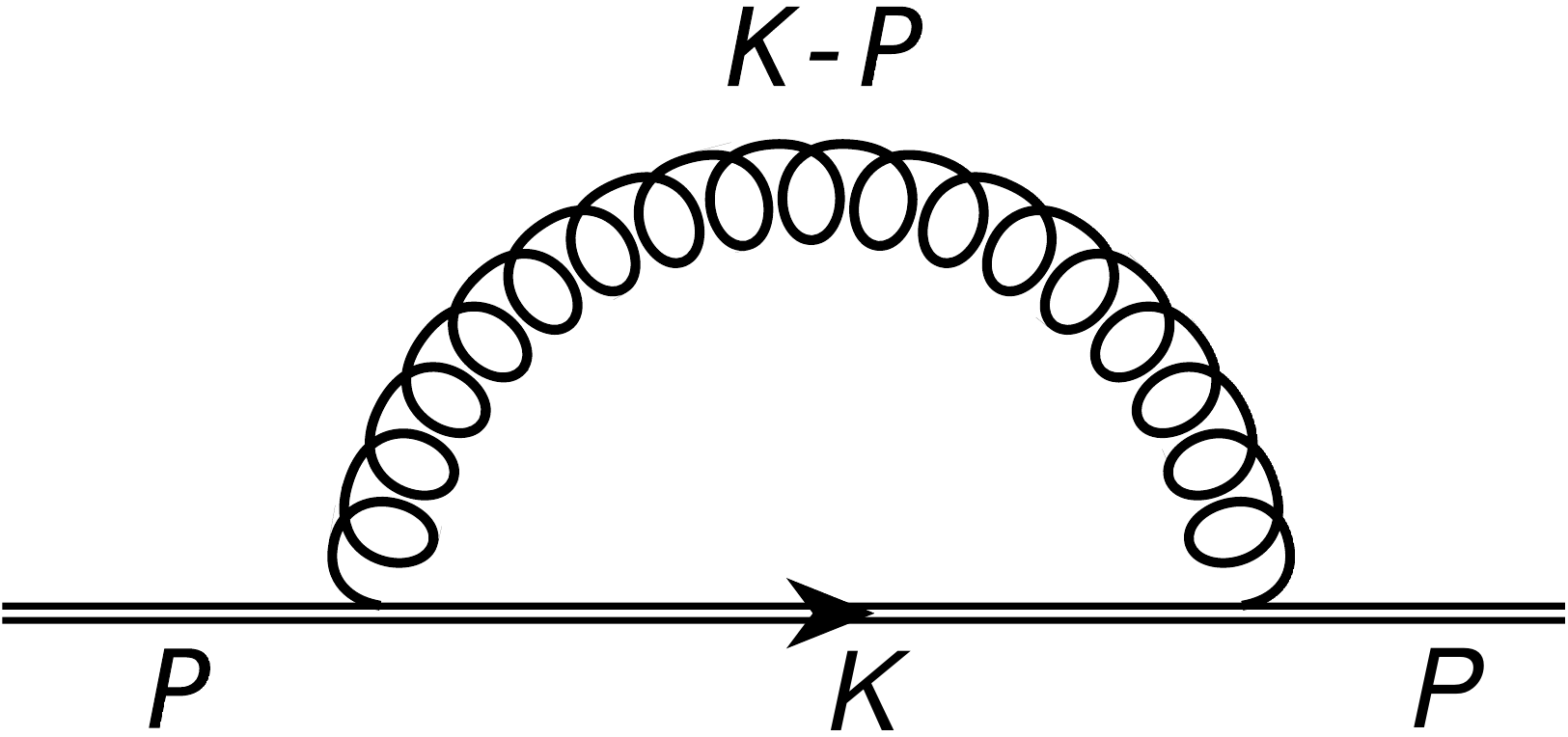} 
			\caption{Self-energy diagram for a quark in presence of background magnetic field. 
				The double line indicates the modified quark propagator in presence of weak magnetic 
				field.}
			\label{quark_se}
		\end{center}
	\end{figure}
\end{center}

Free quark propagator in weak magnetic field is given in Eq.~\eqref{schwinger_prop}.
Now the one loop quark self-energy upto $\mathcal{O}[(q_f B)^2]$ can be written as
\bea
\Sigma(P) &=& g^2 C_F  \sumintf_{\{ K\} } \gm_\mu \bigg(S_0(K)+S_1(K)+S_2(K)\bigg) \gm^\mu
 \frac{1}{(K-P)^2}\nn\\
 &=& \Sigma_0+\Sigma_1+\Sigma_2. 
\eea
From \eqref{a_calc} structure function $a$ can be written upto $\mathcal{O}[(q_f B)^2]$ as 
\bea
a&=& \frac{1}{4}\frac{\Tr(\Sigma_0 \, \slashed P)-(P\cdot u) \Tr(\Sigma_0\, \slashed u)}{(P\cdot u)^2-P^2}+\frac{1}{4}\frac{\Tr(\Sigma_2 \, \slashed P)-(P\cdot u) \Tr(\Sigma_2\, \slashed u)}{(P\cdot u)^2-P^2},\\
&=& a_0+a_B,
\eea
where $a_0$ is purely thermal contribution $(B=0)$ and $a_B$ is the magnetic correction of $\mathcal{O}[(q_f B)^2]$ coming from $\Sigma_2$.
The  $\mathcal{O}[(q_f B)]$ corrections coming from $\Sigma_1$ vanish due to the trace of odd number of gamma matrices.

Similarly structure function $b$ can be written as,
\bea
b&=& \frac{1}{4}\frac{-(P\cdot u)\Tr(\Sigma_0 \, \slashed P)+P^2 \Tr(\Sigma_0\, \slashed u)}{(P\cdot u)^2-P^2}+ \frac{1}{4}\frac{-(P\cdot u)\Tr(\Sigma_2 \, \slashed P)+P^2 \Tr(\Sigma_2\, \slashed u)}{(P\cdot u)^2-P^2},\\
&=&b_0+b _B.
\eea

The thermal part of the structure functions $a$ and $b$ can be calculated using the quark self energy diagram in Fig.~\ref{quark_se} as~\cite{Das:2017vfh}
\bea
a_0(p_0,p)&=&-\frac{m_{\mathrm{th}}^2}{p^2}\int \frac{d\Omega}{4\pi} \frac{p\cdot \hat k}{P\cdot \hat K},\\
\label{struc_a}
b_0(p_0,p)&=& \frac{m_{\mathrm{th}}^2}{p}\int \frac{d\Omega}{4\pi} \frac{(P\cdot u)(p\cdot \hat k)-p^2}{P\cdot \hat K},
\eea
where thermal mass is given as
\bea
m_{\mathrm{th}}^2&=&\frac{g^2C_F T^2}{8}\Big( 1+4\hat \mu^2\Big)
\eea
with $\hat \mu=\mu/2\pi T$ and $C_F=(N_c^2-1)/2 N_c$.


Now we derive the  $\mathcal{O}[(q_f B)^2]$ corrections to the structure functions $a$ and $b$. To get the expression of $a_B$ and $b_B$ we need to perform the following sum-integrations.
\bea
\Tr\big({\Sigma_2 \,\slashed u}\big)&=& 16 g^2 C_F (q_f B)^2 T\sumintf_{\{ K\} } \frac{k_\perp^2 k_0}{(K^2-m_f^2)^4 \,Q^2}\nn\\
&=& \frac{16}{6}g^2 C_F (q_f B)^2 \int \frac{k^2\, dk\, d\Omega }{(2\pi)^3} \frac{\partial^3}{\partial (m_f^2)^3}\bigg[-\frac{k_\perp^2}{4 \sqrt{k^2+m_f^2}}(n_F( \sqrt{k^2+m_f^2})+n_B( \sqrt{k^2+m_f^2}))\bigg]\nn\\
&&\times \bigg(\frac{1}{p_0-p\cdot \hat k}+\frac{1}{p_0+p\cdot \hat k}\bigg)\nn\\
&=&\frac{g^2 C_F (q_f B)^2}{6 \pi^3 T^2}\Gamma(5)  \frac{\partial^3}{\partial (y^2)^3}\bigg[h_5(y)+f_5(y)\bigg]\int d\Omega \frac{\hat k_\perp^2}{P\cdot \hat K}\nn\\
&=& \frac{g^2 C_F T (q_f B)^2}{8 \pi m_f^3}\int \frac{d \Omega}{4\pi} \frac{\hat k_\perp^2}{P\cdot \hat K},
\eea
where we have used well-known functions 
\bea
f_{n+1}(y)= \frac{1}{\Gamma(n+1)} \int_0^{\infty} \frac{dx \, x^n}{\sqrt{x^2+y^2}}\, n_F\big(\sqrt{x^2+y^2}\big)
\eea
and 
\bea
h_{n+1}(y)= \frac{1}{\Gamma(n+1)} \int_0^{\infty} \frac{dx \, x^n}{\sqrt{x^2+y^2}}\, n_B\big(\sqrt{x^2+y^2}\big),
\eea
 which satisfy the recursion relations~\cite{Bandyopadhyay:2017cle},
 \bea
 \frac{\partial f_{n+1}}{\partial y^2}&=& -\frac{f_{n-1}}{2n},\\
  \frac{\partial h_{n+1}}{\partial y^2}&=& -\frac{h_{n-1}}{2n}.
 \eea
 Expressions for $f_1(y)$ and $h_1(y)$ are given as
 \bea
 f_1(y)&=& \frac{\pi}{2y}+\frac{1}{2} \ln\left(\frac{y}{4 \pi}\right)+\, \dotsi ,\nn\\
 h_1(y)&=&-\frac{1}{2} \ln\left(\frac{y}{4 \pi}\right)+\frac{1}{4} \aleph(z)+ \, \dotsi \,,
 \eea 
  where $\aleph(z)$ is defined in Eq.~\eqref{aleph_ue}.
  
 Now we compute the following trace as
\bea
\Tr\big({\Sigma_2 \,\slashed P}\big)&=& X_1+X_2+X_3+X_4,
\eea
where 
\bea
X_1&=&-16 g^2 C_F (q_f B)^2 \sumintf_{\{ K\} } \frac{k \,(p.\hat k)}{(K^2-m_f^2)^3 Q^2}\nn\\
&=&-8 g^2 C_F (q_f B)^2 \int \frac{k^2\, dk\, d\Omega }{(2\pi)^3} \frac{\partial^2}{\partial (m_f^2)^2}\bigg[\frac{k \,(p.\hat k)}{4\, q \sqrt{k^2+m_f^2}}(n_F( \sqrt{k^2+m_f^2})+n_B( \sqrt{k^2+m_f^2}))\bigg]\nn\\
&&\times \bigg(\frac{1}{p_0+p\cdot \hat k}-\frac{1}{p_0-p\cdot \hat k}\bigg)\nn\\
&=&-\frac{g^2 C_F (q_f B)^2}{8 \pi^3 T^2}  \frac{\partial}{\partial y^2}\bigg[h_1(y)+f_1(y)\bigg]\int d\Omega \frac{p \cdot \hat k}{P\cdot \hat K}\nn\\
&=& \frac{g^2 C_F T (q_f B)^2}{8 \pi m_f^3}\int \frac{d \Omega}{4\pi} \frac{p \cdot \hat k}{P\cdot \hat K}.
\eea
By performing similar calculations we get,
\bea
X_2&=& 16 g^2 C_F (q_f B)^2 \sumintf_{\{ K\} } \frac{k \,\hat k_3\, p_3}{(K^2-m_f^2)^3 Q^2}\nn\\
&=&-\frac{g^2 C_F T (q_f B)^2}{8 \pi m_f^3}\int \frac{d \Omega}{4\pi} \frac{p_3 \,\hat k_3\,}{P\cdot \hat K},\\
X_3&=& 16 g^2 C_F (q_f B)^2 \sumintf_{\{ K\} } \frac{k_\perp^2 \, k_0\, p_0}{(K^2-m_f^2)^4 Q^2}\nn\\
&=&\frac{g^2 C_F T (q_f B)^2}{8 \pi m_f^3}\int \frac{d \Omega}{4\pi} \frac{p_0\, \hat k_\perp^2 }{P\cdot \hat K},\\
X_4&=&- 16 g^2 C_F (q_f B)^2 \sumintf_{\{ K\} } \frac{k_\perp^2 \, k(p\cdot \hat k)}{(K^2-m_f^2)^4 Q^2}\nn\\
&=&-\frac{g^2 C_F T (q_f B)^2}{8 \pi m_f^3}\int \frac{d \Omega}{4\pi} \frac{\,\hat k_\perp^2 (p\cdot \hat k) }{P\cdot \hat K}.
\eea
So we obtain the expression of $a_B$ and $b_B$ as 
\bea
a_B&=& \frac{g^2 C_F T (q_f B)^2}{32 \pi m_f^3} \frac{1}{p^2}\int \frac{d \Omega}{4\pi} \,\frac{\hat k_3^2\, (p\cdot \hat k) -p_3 \hat k_3}{P\cdot \hat K},\nn\\
b_B &=& \frac{g^2 C_F T (q_f B)^2}{32 \pi m_f^3} \frac{1}{p}\int \frac{d \Omega}{4\pi} \, \,\frac{\hat k_3(P\cdot u)(\hat p_3\,-\hat p\cdot \hat k\, \hat k_3)-p\, \hat k_\perp^2}{P\cdot \hat K}.
\eea

Other two structure functions are given as \cite{Das:2017vfh}
\bea
c'&=& -m_{\text{eff}}^2 \int \frac{d\Omega}{4\pi} \frac{\hat K\cdot n}{P\cdot \hat K},\nn\\
d'&=&m_{\text{eff}}^2 \int \frac{d\Omega}{4\pi} \frac{\hat K\cdot u}{P\cdot \hat K},
\eea
where the contribution comes only from $\Sigma_1$ term. The contributions from $\Sigma_0$ and $\Sigma_2$ vanish due the trace of odd number of gamma matrices.

Here we note that
\bea
m_{\text{eff}}^2 &=& 4 g^2\,C_F\, M_f^2(T,m_f,q_f B),
\eea
where the thermomagnetic mass for flavor $f$ is given as
\bea
M_f^2(T,m_f,q_fB)&=&\frac{q_fB}{16\pi^2}\bigg[-\frac{1}{4}\aleph(z)-\frac{\pi T}{2 m_f}-\frac{\gamma_E}{2}\bigg] ,
\label{mag_mass}
\eea
where $\aleph(z)$ is defined in Eq.~\eqref{aleph}.

We rearrange the inverse of effective propagator in different way,
\bea
S_{\text {eff}}^{-1}(P)&=& (1+a)\, \slashed{P}+b \,\slashed u +c'\, \gm_5\, \slashed u+ d' \, \gm_5\, \slashed n-m_f\,  \mathbb{I}\nn\\
&=& c\, p_0 \,\gm_0 - d p_i\,\gm_i+c'\, \gm_5\, \slashed u+ d' \, \gm_5\, \slashed n-m_f\,  \mathbb{I},
\eea
where 
\bea
c&=&1+(a_0+\frac{b_0}{p_0})+(a_B+\frac{b_B}{p_0})=1-a'_0-a'_B, \nn\\
d&=&1+a_0+a_B,
\label{candd}
\eea
with
\bea
a_0&=& \frac{m_{\text{th}}^2}{p^2} (1-\mathcal{T}_p),\nn\\
a'_0&=& \frac{m_{\text{th}}^2}{p_0^2} \mathcal{T}_p, \nn\\
a_B&=& \frac{m'^2_{\text {eff}}}{6 p^2}\bigg(1-\frac{3\, p_3^2}{p^2}\bigg)\bigg(\frac{3\, p_0^2}{p^2}(1-\mathcal{T}_p)-2+3 \mathcal{T}_p\bigg) ,\nn\\
a'_B&=& \frac{m'^2_{\text {eff}}}{2 p_0^2}\bigg(3\, \frac{p_0^2\,p_3^2}{p^4}(1-\mathcal{T}_p)+\mathcal{T}_p+\frac{1}{p^2}\bigg(p_3^2\,\mathcal{T}_p-p_0^2(1-\mathcal{T}_p)\bigg)\bigg),\nn\\
c'&=& \frac{p_3\,m_{\text{eff}}^2}{p^2} (1-\mathcal{T}_p),\nn\\
d'&=& \frac{m_{\text{eff}}^2}{p_0} \mathcal{T}_p.
\eea
We expressed all the structure functions in terms of
\bea
\mathcal{T}_p &=& \int \frac{d\Omega}{4 \pi} \frac{p_0}{p_0-p\cdot \hat k}.
\label{ang_avg}
\eea

Here we have defined 
\bea
m'^2_{\text {eff}}&=& \frac{g^2 C_F (q_f B)^2 T}{32 \,\pi \,m_f^3}
\eea
with $C_F=(N_c^2-1)/2 N_c$.
We can see that $m'^2_{\text {eff}}$ and $m_{\text{eff}}^2$ diverge when current quark mass vanishes ($m_f \rightarrow 0$). It is regulated by thermal mass $m_{th}$ of the fermion as discussed in Refs.~\cite{Kapusta:2006pm,Ayala:2014uua}.


\section{Sum-integrals}
\label{sum_int}

The dimensionally regularized sum integrals are defined as
\bea
\sumintf_{\{ P\} } &=& \left(\frac{e^{\gamma_E}\Lambda^2}{4\pi}\right)^\epsilon 
T\sum\limits_{\substack{p_0=(2n+1)\pi 
T i +\mu}}\int\frac{d^{d-2\epsilon}p}{(2\pi)^{d-2\epsilon}},
\eea
where $\Lambda$ can be identified as the $\overline{MS}$ renormalization scale which also 
introduces the factor $\left(\frac{e^{\gamma_E}}{4\pi}\right)^\epsilon$ 
along with it where $\gamma_E$ is the Euler-Mascheroni constant. 

The
sum integrals are related by the following equations.   
\bea
\sumintf_{\{ P\} }\frac{1}{P^4}= -\frac{d-2}{2}\sumintf_{\{ P\} 
}\frac{1}{p^2P^2},\\
\sumintf_{\{ P\} }\frac{1}{p^2P^4}= -\frac{d-4}{2}\sumintf_{\{ P\} 
}\frac{1}{p^4P^2}.
\eea
\subsection{One-loop sum integrals }
\label{suminti_th}
We list the fermionic sum-integrals as~\cite{Bandyopadhyay:2017cle,Chakraborty:2002yt}

\bea
\sumintf_{\{P\}}\frac{1}{P^2}
&=&\frac{T^2}{24}\left(\frac{\Lambda}{4\pi 
T}\right)^{2\epsilon}\left[1+12\hat\mu^2\right],\\
\sumintf_{\{ P\} 
}\frac{1}{P^4}&=&\frac{1}{\left(4\pi\right)^2}\left(\frac{\Lambda}{4\pi 
T}\right)^{2\epsilon}\Bigg[\frac{1}
{\epsilon}-\aleph(z)\Bigg],\\
\sumintf_{\{ P\} 
}\frac{p_3^2}{p^2 P^4}&=&\frac{1}{3 \left(4\pi\right)^2}\left(\frac{\Lambda}{4\pi 
	T}\right)^{2\epsilon}\Bigg[\frac{1}
{\epsilon}+\frac{2}{3}-\aleph(z)\Bigg],\\
\sumintf_{\{P\}}\frac{1}{P^6}&=&
\frac{1}{(2\pi)^4}\left(\frac{\Lambda}{4\pi T}\right)^{2\epsilon}\frac{\gimel(z)}{32T^2}, \\
\sumintf_{\{P\}}\frac{1}{p^4P^2}&=&-\frac{1}{(2\pi)^4}\left(\frac{\Lambda}
{4\pi T}\right)^{2\epsilon} \frac{\gimel(z)}{4T^2},\\
\sumintf_{\{P\}}\frac{1}{p^2P^4}&=&-\frac{1}{(2\pi)^4}\left(\frac{\Lambda}
{4\pi T}\right)^{2\epsilon}\frac{\gimel(z)}{8T^2}, \\
\sumintf_{\{P\}}\frac{p_3^2}{p^2P^6} &=& \frac{1}{(2\pi)^4}\left(\frac{\Lambda}
{4\pi T}\right)^{2\epsilon} \frac{\gimel(z)}{96T^2}, \\
\sumintf_{\{P\}}\frac{p_3^2}{p^4P^4}
&=&-\frac{1}{(2\pi)^4}\left(\frac{\Lambda}
{4\pi T}\right)^{2\epsilon} \frac{\gimel(z)}{24T^2}.
\eea
Here we list the frequently used functions in the sum-integrals
\bea
\aleph(z)&\equiv& \Psi(z)+\Psi(z^*),\label{aleph_ue}\\
\gimel(z)&\equiv&\frac{d^2}{dz^2}\Big( \Psi(z)+\Psi(z^*) \Big),
\eea
where $z$ is a general complex number, here $z=1/2-i \hat \mu$. $\zeta$ and $\Psi$ denote the Riemann Zeta function and the digamma function respectively. The digamma function can be written as
\bea
\Psi(z)\equiv\frac{\Gamma'(z)}{\Gamma(z)}.
\eea
We write the functional form of $\aleph(1/2-i\hat \mu)$ and $\gimel(1/2-i\hat \mu)$ for small $\hat\mu$ below. 
\bea
 \aleph(z)&=&-2\gamma_E-4\ln 
2+14\zeta(3)\hat{\mu}^2-62\zeta(5)\hat{\mu}^4+254\zeta(7)\hat{\mu}^6+{\cal 
O}(\hat{\mu}^8),\label{aleph}\\
\gimel(z)&=&-4\bigg[7\zeta(3)-186\zeta(5) \hat\mu^2+1905\zeta(7) \hat\mu^4-14308 \zeta(9) \hat\mu^6\bigg]+ \mathcal O(\hat \mu^8).\label{alpha}
\eea
\subsection{One-loop HTL sum integrals used in the magnetic case}
\label{htl_sum_inti}
We also need  one-loop HTL sum integrals which involve the angular average defined earlier in Eq.~\eqref{ang_avg}. For brevity, henceforth we will use the notation $c=\cos\theta$ for single angular average and $c_i=\cos\theta_i$ for multiple angular averages. We list the sum integrals below:

\bea
\sumintf_{\{P\}}\frac{1}{P^4}\mathcal T_p&=&\frac{d-4}{d-5}\sumintf_{\{P\}}\frac{1}{P^4}\\
\sumintf_{\{P\}}\frac{1}{p^2P^4}\mathcal T_p&=&\Big(\Delta_1-\frac{d-4}{2}\Delta_2\Big) \sumintf_{\{P\}}\frac{1}{p^4P^2},, \\
\sumintf_{\{P\}}\frac{p_3^2}{p^2 P^4}\mathcal T_p&=&\bigg(\frac{d-4}{d-2}(1+\Delta_0)+\frac{2}{d-2}\Delta_3'\bigg)\Delta_3\sumintf_{\{P\}}\frac{1}{P^4}\\
\sumintf_{\{P\}}\frac{\mathcal T_p^2}{p^2p_0^2P^2}&=& \Delta_4\sumintf_{\{P\}}\frac{1}{p^4P^2},,\\
\sumintf_{\{P\}}\frac{p_3^2}{p^4P^4} \mathcal T_p
&=&\Big(\Delta_1-\frac{d-4}{2}\Delta_2\Big) \Delta_3\sumintf_{\{P\}}\frac{1}{p^4P^2},\\
\sumintf_{\{P\}}\frac{1}{p_0^2P^4} \mathcal T_p^2
&=&\Delta_5\sumintf_{\{P\}}\frac{1}{p^4P^2},\\
\sumintf_{\{P\}}\frac{p_3^2 }{p^4P^4}\mathcal T_p^2 &=&\Delta_3 \Delta_6 
\sumintf_{\{P\}}\frac{1}{p^4P^2},\\
\sumintf_{\{P\}}\frac{p_3^2}{p^2p_0^2P^4}  \mathcal T_p^2
&=&\Delta_3 \Delta_5\sumintf_{\{P\}}\frac{1}{p^4P^2},
\label{fermion_sum_int1}
\eea
where $\Delta$'s are the various angular averages which we list below.
 The symbol $\langle\rangle_c$ in the angular averages depicts the standard definition used in Ref.~\cite{Andersen:2002ey}.
\bea
\Delta_0 &=& \left\langle \frac{c^2}{1-c^2}\right\rangle_c=-\frac{1}{2\eps}+\mathcal{O}[\epsilon]^3,\\
\Delta_1 &=& \left\langle \frac{c^{6-d}-c^2}{(1-c^2)^2}\right\rangle_c = 
\frac{1}{4\eps}-\frac{1}{4}+\ln 2+\eps\bigg[ -\frac{3}{4}+\frac{\pi^2}{6}+(\ln 2)^2-2\ln 2\bigg]+\mathcal{O}[\epsilon]^2 ,\\
\Delta_2 &=& \left\langle \frac{1}{1-c^2}\right\rangle_c=-\frac{1}{2\eps}+1+\mathcal{O}[\epsilon],\\
\Delta_3&=& \left\langle c^2\right\rangle_c = \frac{1}{3}+\frac{2\eps}{9}+\mathcal{O}[\epsilon]^2,\\
\Delta_3' &=& \left\langle \frac{1-c^{4-d}}{(1-c^2)^2}\right\rangle_c=-\frac{1}{4\eps}+\frac{1}{4}+\frac{3\eps}{4}-\frac{3\eps^2}{4}+\mathcal{O}[\epsilon]^3,\\
\Delta_4&=& \left\langle \frac{1-c_1^{6-d}}{(c_1^2-c_2^2)(1-c_1^2)}+c_1\lrarrow c_2\right\rangle_{c_1,c_2}=\frac{1}{12}\bigg[24\ln 2 -\pi^2 \bigg]+\mathcal{O}[\epsilon] ,\\
\Delta_5&=& \left\langle\frac{c_1^{6-d}-c_1^2}{(c_1^2-c_2^2)(1-c_1^2)^2}-\frac{d-4}{2}\frac{1}{(c_1^2-c_2^2)(1-c_1^2)}+c_1\lrarrow c_2\right\rangle_{c_1,c_2}  = \frac{1}{24} 
  \bigg[ 6-\pi^2\bigg] +\mathcal{O}[\epsilon] ,\\
\Delta_6 &=& \left\langle\frac{1-2c_1^2+c_1^{8-d}}{(c_1^2-c_2^2)(1-c_1^2)^2}-\frac{d-4}{2}\frac{1}{(c_1^2-c_2^2)(1-c_1^2)}+c_1\lrarrow c_2\right\rangle_{c_1,c_2}  = \frac{1}{8}\bigg[ 2-\pi^2+16\ln 2\bigg]+\mathcal{O}[\epsilon].\nn\\
\label{ang_int}
\eea
Using the expressions of angular averages we obtain the results of HTL sum integrals for magnetic case as
\bea
\sumintf_{\{ P\} 
}\frac{1}{P^4}\mathcal{T}_p&=&\frac{1}{2\left(4\pi\right)^2}\left(\frac{\Lambda}{4\pi 
	T}\right)^{2\epsilon}\Bigg[\frac{1}
{\epsilon}+1-\aleph(z)\Bigg],\\
\sumintf_{\{P\}}\frac{1}{p^2P^4}\mathcal T_p&=&-\frac{1}{(2\pi)^4}\left(\frac{
	\Lambda}{4\pi T}\right)^{2\epsilon}\frac{\gimel(z)}{16T^2}\Big(-1+4\ln 2 \Big), \\
\sumintf_{\{ P\} 
}\frac{p_3^2}{p^2 P^4}\mathcal{T}_p&=&\frac{1}{ 6 \left(4\pi\right)^2}\left(\frac{\Lambda}{4\pi 
	T}\right)^{2\epsilon}\Bigg[\frac{1}
{\epsilon}+\frac{5}{3}-\aleph(z)\Bigg],\\
\sumintf_{\{P\}}\frac{1}{p^2p_0^2P^2}\mathcal T_p^2&=&\frac{1}{(2\pi)^4}\!\left(\frac{\Lambda}
{4\pi T}\right)^{2\epsilon}\frac{\gimel(z)}{48T^2}\Big(\pi^2-24\ln 2\Big),\\
\sumintf_{\{P\}}\frac{p_3^2}{p^4P^4} \mathcal T_p&=&-\frac{1}{(2\pi)^4}\left(\frac{\Lambda}
{4\pi T}\right)^{2\epsilon} \frac{\gimel(z)}{48T^2}\Big(-1+4\ln 2 \Big),\\
\sumintf_{\{P\}}\frac{1}{p_0^2P^4} \mathcal T_p^2
&=&\frac{1}{(2\pi)^4}\left(\frac{\Lambda}
{4\pi T}\right)^{2\epsilon} \frac{\gimel(z)}{96T^2}\Big( \pi^2-6\Big),\\
\sumintf_{\{P\}}\frac{p_3^2 }{p^4P^4}\mathcal T_p^2 &=&\frac{1}{(2\pi)^4}\left(\frac{\Lambda}
{4\pi T}\right)^{2\epsilon} \frac{\gimel(z)}{96T^2}\Big( -2+\pi^2-16\ln 2\Big),\\
\sumintf_{\{P\}}\frac{p_3^2 }{p^2p_0^2P^4} \mathcal T_p^2
&=&\frac{1}{(2\pi)^4}\left(\frac{\Lambda}
{4\pi T}\right)^{2\epsilon} \frac{\gimel(z)}{288   T^2}\Big( \pi^2-6\Big).
\eea

\end{document}